\documentclass[12pt,epsfig]{article}
\usepackage{graphicx}
\usepackage{epsfig}

\title{X-Radiography of Cargo Containers} 

\author
{J. I. Katz,$^{1,2,\ast}$ G. S. Blanpied,$^{1,3}$
K. N. Borozdin,$^1$ C. Morris$^1$\\
\\
\normalsize{$^1$Los Alamos National Laboratory, Los Alamos, N. Mex. 87545,
USA}\\
\normalsize{$^2$Department of Physics and McDonnell Center for the Space
Sciences}\\
\normalsize{Washington University, St. Louis, Mo. 63130 USA}\\
\normalsize{$^3$Department of Physics, University of South Carolina,
Columbia, S. C. 29208 USA}\\
\\
\normalsize{$^\ast$Corresponding author: J. I. Katz}\\
\normalsize{Dept. Physics, Washington University, St. Louis, Mo. 63130}\\
\normalsize{tel: 314-935-6202, facs: 314-935-6219, email:
katz@wuphys.wustl.edu.}
}

\date{}

\begin{document} 


\baselineskip24pt


\maketitle 

\vfill




\newpage

\begin{abstract}
The problem of detecting a nuclear weapon smuggled in an
ocean-going cargo container has not been solved, and the detonation of such
a device in a large city could produce casualties and property damage
exceeding those of September 11, 2001 by orders of magnitude.  Any means
of detecting such threats must be fast and cheap enough to screen the
millions of containers shipped each year, and must be capable
of distinguishing a threatening quantity of fissionable material from the
complex loading of masses of innocent material found in many containers.
Here we show that radiography with energetic X-rays produced by a 10 MeV
electron accelerator, taking advantage of the high density and specific
atomic properties of fissionable material, may be a practical solution.
\end{abstract}

\section{Introduction}

Approximately 7,000,000 cargo containers enter the United States by sea
each year, and about 9,000,000 by land$^1$.  Roughly comparable numbers are
shipped between other countries.  These containers, only a comparatively
few of which are opened for inspection$^2$, offer a terrorist a potential
means of smuggling a nuclear weapon across international borders.  Twice in
recent years fifteen pound (7 kg) chunks of depleted uranium, harmless
itself but massive enough to resemble threatening quantities of
weapons-grade uranium or plutonium, are known to have passed border
inspection without detection$^{3,4}$.  In this paper we present the results
of Monte Carlo calculations showing that radiographs taken with
sufficiently energetic X-rays are capable of detecting threatening
quantities of fissionable material, even in a container loaded with other
massive absorbers in a complex geometry.

X-ray radiography is the traditional method of looking inside opaque
objects$^5$.  It works very well for comparatively small objects, but
the dimensions (2.6 m $\times$ 2.6 m $\times$ 12 m) and heavy and spatially
complex loading of the standard 40-foot cargo container present serious
obstacles.  At a mean density of 0.3 g cm$^{-3}$ (this 24 MT [metric ton]
load is typical, although loads up to 30 MT are permitted) its column
density across its shortest dimension is 78 gm cm$^{-2}$.  The
scattering of X-rays of energies less than a few hundred KeV is well
described by the Thomson cross-section$^6$, giving an opacity of about
0.2 cm$^{2}$/g for most materials.  This leads to 15.6 e-folds (a factor of
$1.7 \times 10^{-7}$) of beam attenuation, which precludes use of these
lower energy X-rays.

Fortunately, at higher energies the scattering cross-section is described by
the Klein-Nishina formula$^6$, and declines nearly as the reciprocal
of the energy.  For high-Z materials such as uranium and plutonium another
absorption process, electron-positron pair production, whose cross-section
increases with energy, dominates the attenuation above about 3 MeV$^{6,7}$.
Pair production is less important for lower-Z materials, so their
opacities flatten out or continue to decrease as the energy increases, as
shown in Figure 1.

The beam attenuation across a container filled with 0.3 g cm$^{-3}$ of
low or medium-Z material is then only about 2 e-folds (a factor of 0.14) at
energies of several MeV,
so that X-ray radiography becomes possible.  Further, because the opacity
(in cm$^2$/g) is larger for high-Z materials, they will stand out even
more strongly in radiographs than indicated by their high density alone.

\section{Calculations}

The multiple physical processes and complex geometries required to model
X-ray radiography imply that quantitative results can only be obtained
from Monte Carlo calculations.  It is necessary to include electron and
positron elastic scattering, bremsstrahlung, collisional ionization and
Coulomb pair production, pair annihilation, photon Compton and coherent
scattering, photoionization and photopair production and radiative
recombination.  The spatial, angular and energy distribution of photons,
electrons and positrons must be tracked.  In auxiliary calculations
photoneutron processes and neutron transport and capture must be calculated
as well.  In order to handle these computationally formidable tasks we used
the MCNPX code$^{8-10}$.

We first consider a 5 kg sphere of $\delta$-plutonium ($r = 4.22$ cm) at
the center of a container otherwise uniformly filled with iron to a density
of 0.3 g cm$^{-3}$.  The X-ray source is a beam of 10 MeV electrons that
radiate bremsstrahlung when stopped by a 7 mm thick tungsten converter slab
at a height of 5.2 m above the top of the container (the height enables a
single X-ray source to illuminate the entire container width).  The
converter also serves as a high-pass spectral filter for the emitted
radiation.

Extensive collimation is necessary to reduce the scattering of radiation
into the deep absorption minimum produced by the plutonium sphere.  Below 
the converter there is a 1.1 cm wide slot collimator made of tungsten 10 cm
thick.  A similar slot collimator above the container matches a 1 cm wide
detector array.  The detectors are modeled as a transverse row of point
sensors 20 cm below the container, spaced 1 cm apart, which respond to the
X-ray energy flux, a fair approximation to the behavior of several practical
scintillators.  A final Bucky$^{11}$ collimator between the container
and the detectors consists of a 16 cm thick slab of tungsten with holes of
0.5 cm diameter bored along the lines from each detector to the radiation
source.  The incident electron beam is taken to be $13^\circ$ from vertical.
The geometry is shown in Figure 2.

\section{Results}

The discriminating power of high energy X-ray radiography is demonstrated
by Figure 3, in which the plutonium sphere is clearly and unambiguously
revealed.  The statistical uncertainty in the results may be estimated from
the point-to-point fluctuations in the signal, and is $< 10\%$.  The entire
length of a 40-foot (12 m) cargo container may be scanned with 1200
exposures as it is continuously moved through a pulsed X-ray beam.  MeV
electron accelerators may produce micro-second pulses at a rate of several
hundred per second, so the required scanning time is only a few
seconds$^{12}$.

Many containers will contain bodies of innocent dense medium-Z material
(large castings such as engine blocks, ingots, rod stock, {\it etc.\/}), and
a terrorist may fill the empty space in his container with such objects in
order to disguise a dense piece of fissionable material.  Radiography must
identify, or exclude the presence of, a threat in such a cluttered
environment.   Figure 4 therefore shows the radiograph of the same sphere of
plutonium at the center of a very cluttered container (Figure 2).  In
addition to the threat object, it contains 230 spheres of half-density iron
(a model of an automotive engine block, allowing for internal voids), each
20 cm in radius, totaling 30 MT.  The iron spheres are in square arrays of
50 cm spacing, in planes 0.55 m and 1.05 m below the container's midplane.

\section{Discussion}

If the direction of irradiation were vertical the plutonium sphere would not
be detectable because the line of sight through it would pass through the
centers of two of the iron spheres, for a total of 314 g cm$^{-2}$ of iron.
It is for this reason that oblique illumination was chosen.  Multiple
oblique angles may be used to reduce further the possibility of concealing
a fissionable threat object behind opaque masses of lower-Z material.  The
plutonium is detectable, even though lines of sight through it also pass
through one of the iron spheres, because its characteristic signature---a
combination of high attentuation and small dimension transverse to the
beam---is found only for massive chunks of high-Z material and for paths
along the long axes of long slender objects.

In innocent cargo long slender dense objects are packed with their longest
axes horizontal, and dense cargoes are spread on the floor of the
container.  Therefore, near-vertical irradiation will only rarely show
regions of intense absorption in innocent cargo.  In contrast, horizontal
irradiation would often find this ``false positive'' result, requiring
manual unloading and inspection.  Another advantage of downward
near-vertical illumination is that the Earth is an effective beam-stop;
combined with a thin lead ground plane, its albedo is negligible and
additional shielding would not be required.

A terrorist could hide his fissionable cargo in the shadow of a very large
and deep absorber (such as a 30 MT cube of solid iron).  Such a threat could
be found by opening the very few containers which show absorption too deep
to see through.  The innocent shipper can avoid a false-positive detection
(and the opening of his container) by ensuring that his cargo not present a
deep, spatially localized, absorption maximum in the known direction of
irradiation.  It is not necessary that radiography find all threats or
exculpate all unthreatening containers, only that it identify all containers
that {\it might\/} contain a threat, and make that number small enough to
permit opening and manual inspection.

There is a premium on using as high energy X-rays (and necessarily high
energy electrons) as possible.  Not only is the overall transmission
increased, but the discrimination between high-Z and low or medium-Z
opacities improves.  In addition, the coherent and Compton scattering
cross-sections are less and the bremsstrahlung radiation pattern and the
Compton scattering cross-section are more forward-peaked$^6$.  Scattered
radiation tends to fill in the deep and spatially localized absorption
minima of chunks of high-Z material, which are their characteristic
signature.  This may be minimized by increasing the electron (and therefore
X-ray) energy, and by use of a Bucky collimator which absorbs scattered
radiation arriving on oblique paths.

The chief objection to the use of more energetic X-rays (and electron
accelerators) is photoneutron production.  For most nuclei the photoneutron
energy threshold is about 8 MeV$^7$, so electron beams of energy
greater than 8 MeV will produce some X-rays energetic enough to make
neutrons and lead to a low level of neutron activation in innocent cargo.
However, at the required intensity of irradiation this is insignificant.
Depositing 10 MeV of X-ray energy (typically about three X-rays) in a 1 cm
$\times$ 1 cm detector on a path through the center of a 5 kg
plutonium sphere in a very cluttered container (Figure 4) will show the
depth of absorption to a factor of about two, sufficient for the image to
show the dense high-Z object.  From the calculated results, this would
require $1.1 \times 10^{11}$ 10 MeV electrons per image slice, or about 0.18
Joule (small compared to the capability of industrial radiographic
accelerators).  The container would be irradiated with about $1.3 \times
10^{-7}$ J/cm$^2$ of X-rays on its upper surface, or a total of about 40 mJ
of energetic X-rays.  Even at photon energies of 10--20 MeV the photoneutron
cross-section is no more than 0.01 of the total cross-section$^7$, so
that these $2.5 \times 10^{10}$ X-rays produce, at most, $2.5 \times 10^8$
photoneutrons.  This should be compared to the cosmic ray neutron production
of 0.1/kg/sec$^{13}$, or $3 \times 10^3$/sec for a 30 MT cargo.  Even the
highest energy radiography produces a neutron fluence and activation less
than that produced by a day of cosmic ray exposure.

The neutron production in the collimators, which absorb nearly all the
X-rays, is also small.  The 1200 pulses required to scan a 40 foot (12 m)
container in 1 cm slices contain $1.3 \times 10^{14}$ electrons.  We have
calculated, again using MCNPX$^{8-10}$, the photoneutron production
in the 7 mm tungsten converter followed by a 10 cm lead collimator.  The
neutron to electron ratio is $7 \times 10^{-6}$ at 10 MeV, $7 \times
10^{-4}$ at 15 MeV and $2.5 \times 10^{-3}$ at 20 MeV (where the
bremsstrahlung spectrum overlaps the nuclear giant dipole resonance$^{14}$).
For 10 MeV electrons the dose to an unshielded operator at 20 m
range who examines one container per minute would be 500 nanoSv/hr (using the
standard relation of flux to dose rate$^{15}$).  This is a factor of 50
times less than the occupational limit of 0.05 Sv/year (25 microSv/hr), and
only a small fraction of the typical 2 mSv/year natural background.  The
advantages of radiography at energies of 10 MeV may be obtained with
acceptable personnel exposure.

\section{Acknowledgements}

We thank R. C. Schirato for pointing out the power of Bucky collimators to
reduce the effects of scattered X-rays in optically thick targets.  This
work was supported by the U. S.  Department of Energy.

\section{References}

\begin{enumerate}
\item U. S. Customs and Border Protection \\
www.cbp.gov/xp/enforcement/international\_activities/csi accessed March 12,
2004.
\item Koonin, S. E. {\it et al.\/} Radiological Warfare (Technical Report
JSR-02-340, MITRE Corp., McLean, Va., 2002).
\item ABC News abcnews.go.com/sections/wnt/DailyNews/sept11\_uranium020911.html
accessed March 12, 2004.
\item ABC News abcnews.go.com/sections/wnt/PrimeTime/sept11\_uranium030910.html
accessed March 12, 2004.
\item {\it Nondestructive Testing Handbook\/} 3rd Ed., V. 4, Bossi, R. H.,
Iddings, F. A., Wheeler, G. C., Moore, P. O. Eds. (Am. Soc. Nondestructive
Testing, Columbus, Ohio, 2002).
\item Bjorken, J. D. and Drell, S. D. {\it Relativistic Quantum Mechanics\/}
(McGraw-Hill, New York, 1964).
\item Los Alamos National Laboratory t2.lanl.gov/data/ndviewer.html accessed
January 12, 2004.
\item Hughes, H. G., Egdorf, H. W., Gallmeier, F. C., Hencricks, J. S.,
Little, R. C., McKinney, G. W., Prael, R. E., Roberts, T. L., Snow, E.,
Waters, L. S. {\it et al.\/} (2002) MCNPX User's Manual Version 2.3.0
(Technical Report LA-UR-02-2607, Los Alamos National Laboratory, Los Alamos,
N. Mex.).
\item Hughes, H. G., Egdorf. H. W., Gallmeier, F. C., Hendricks, J. S.,
Little, R. C., McKinney, G. W., Prael, R. E., Roberts, T. L., Snow, E.,
Waters, L. S. {\it et al.\/} (2002) MCNPX User's Manual Version 2.4.0
(Technical Report LA-CP-02-408, Los Alamos National Laboratory, Los Alamos,
N. Mex.).
\item Hendricks, J. S., McKinney, G. W., Waters, L. S., Roberts, T. L.,
Egdorf, H. W., Finch, J. P., Trellue, H. R., Pitcher, E. J., Mayo, D. R.,
Swinhoe, M. T. {\it et al.\/} (2004) MCNPX Extensions Version 2.5.0
(Technical Report LA-UR-04-0570, Los Alamos National Laboratory, Los Alamos,
N. Mex.).
\item Bucky, G. (1913) A grating diaphragm to cut off secondary rays from the
object {\it Archives of the Roentgen Ray} {\bf 18}, 6--9.
\item {\it BEAMS 2002: 14th International Conference on High Power Particle
Beams,\/} Mehlhorn, T. A., Sweeney, M. A. Eds. (AIP, Melville, N. Y. 2002).
\item Pal, Y. (1967) in {\it Handbook of Physics\/}, eds. Condon, E. U. \&
Odishaw, H. (McGraw-Hill, New York), Figure 11.22.
\item Bohr, A. and Mottelson, B. R. (1969) {\it Nuclear Structure\/}
(Benjamin, New York).
\item Knoll, G. F. (1979) {\it Radiation Detection and Measurement\/}
(Wiley, New York).
\end{enumerate}

\newpage

\section{Figure captions:}

\begin{enumerate}
\item Photoelectric absorption of representative
elements$^7$.  At lower energies the cross-sections (shown per atom)
decrease with increasing energy because of the decline of the Compton
scattering (Klein-Nishina) cross-section, while at higher energies they
increase for high-Z elements (but not for low or medium-Z elements) with
increasing energy because of the increasing pair production cross-section.

\item Radiographing a cargo container.
Schematic diagram (parts not to scale) shows electron beam source,
bremsstrahlung converter, direction of direct X-ray illumination,
collimators, detectors and target geometry used in calculations.  
Propagation of energy on indirect paths as a result of scattering and
absorption followed by reemission is important, and the Bucky collimator
is essential to filter out the scattered radiation, revealing the deep
absorption produced by compact bodies of fissionable material.
Quantitative dimensions are given in the text.

\item Absorption radiograph of a 5 kg plutonium sphere.
This threat object is placed in a 40 foot container otherwise filled with
30 MT of uniformly distributed iron.  The distinctive deep but spatially
localized absorption of dense fissionable material is evident.  Absorption
is defined as the reciprocal of the detected energy in MeV per cm$^2$ per
source electron.  The $x$ and $y$ coordinates are in cm.

\item Absorption radiograph of a 5 kg plutonium sphere in a
cluttered container.  A 5 kg plutonium sphere has been placed in a
container with 30 MT of iron distributed in two planes of half-density
spheres of 40 cm diameter (resembling automotive engine blocks, for
example). The plutonium produces a striking compact absorption peak,
readily distinguishable from the absorption by the other contents of the
container, and identifiable by its combination of small size and deep
absorption.

\end{enumerate}


\newpage

\begin{figure}
\includegraphics[angle=270,width=5.5in]{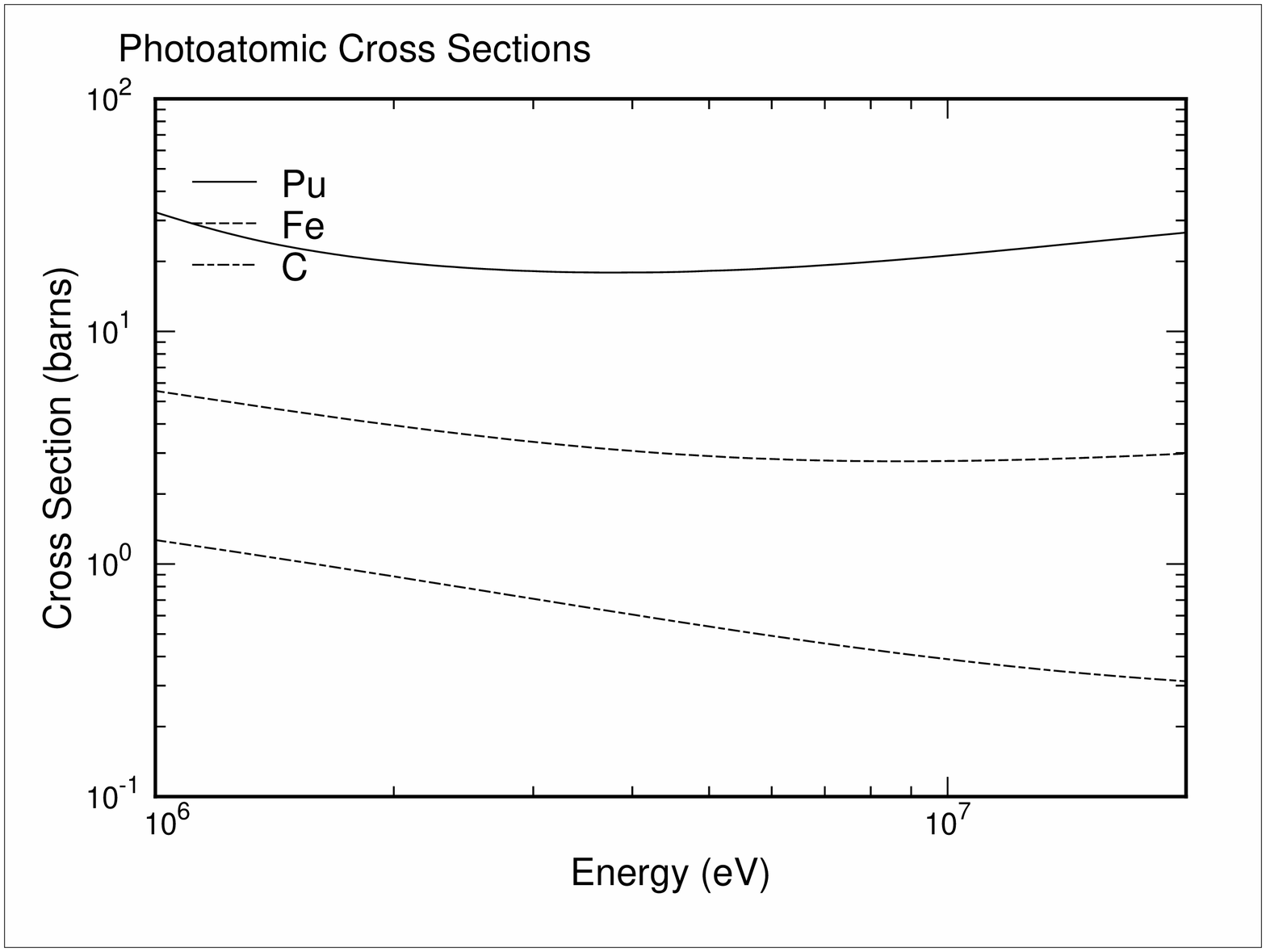}
\end{figure}
\newpage
\begin{figure}
\includegraphics[angle=0,width=5in]{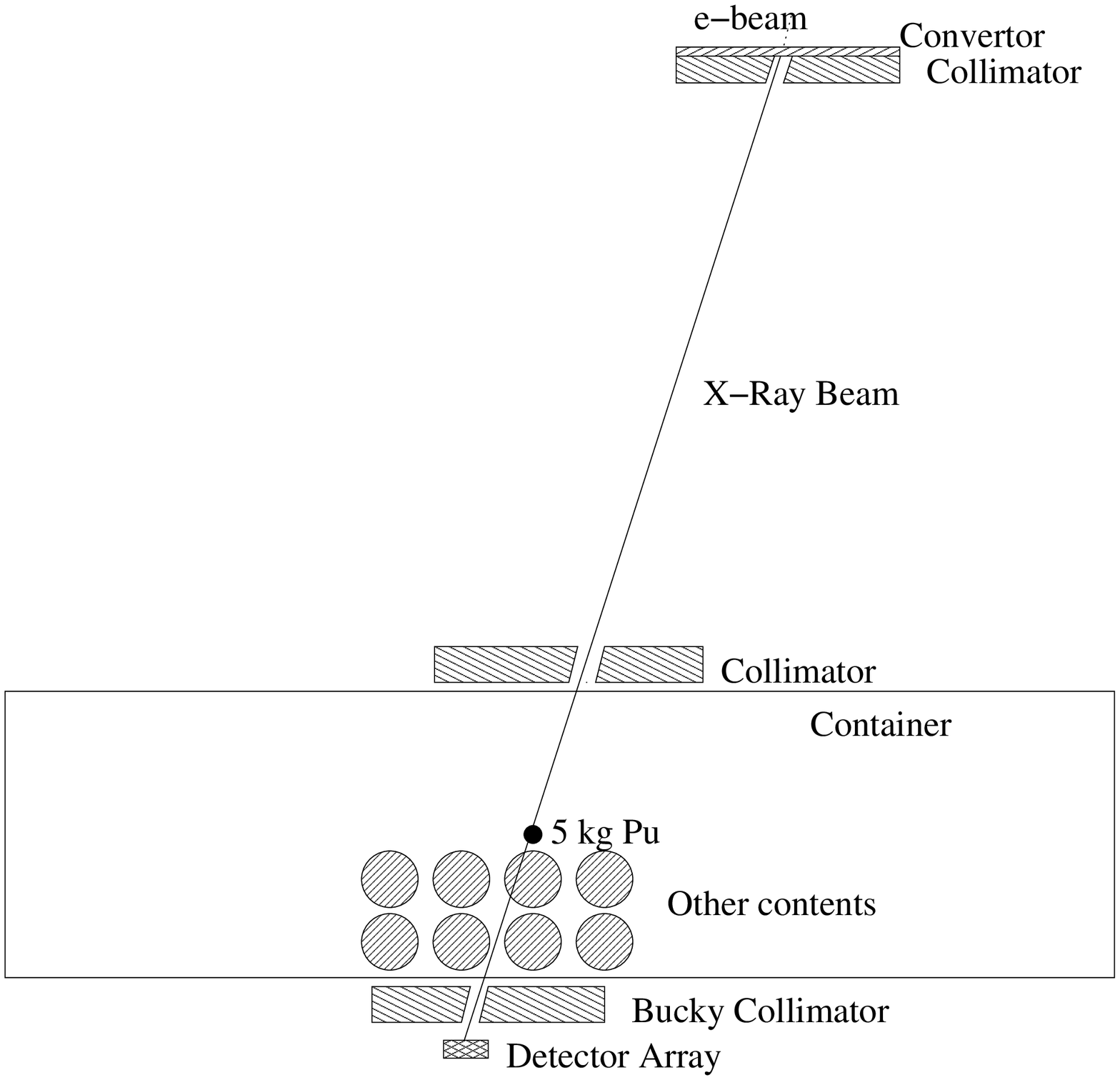}
\end{figure}
\newpage
\begin{figure}
\includegraphics[angle=0,width=5in]{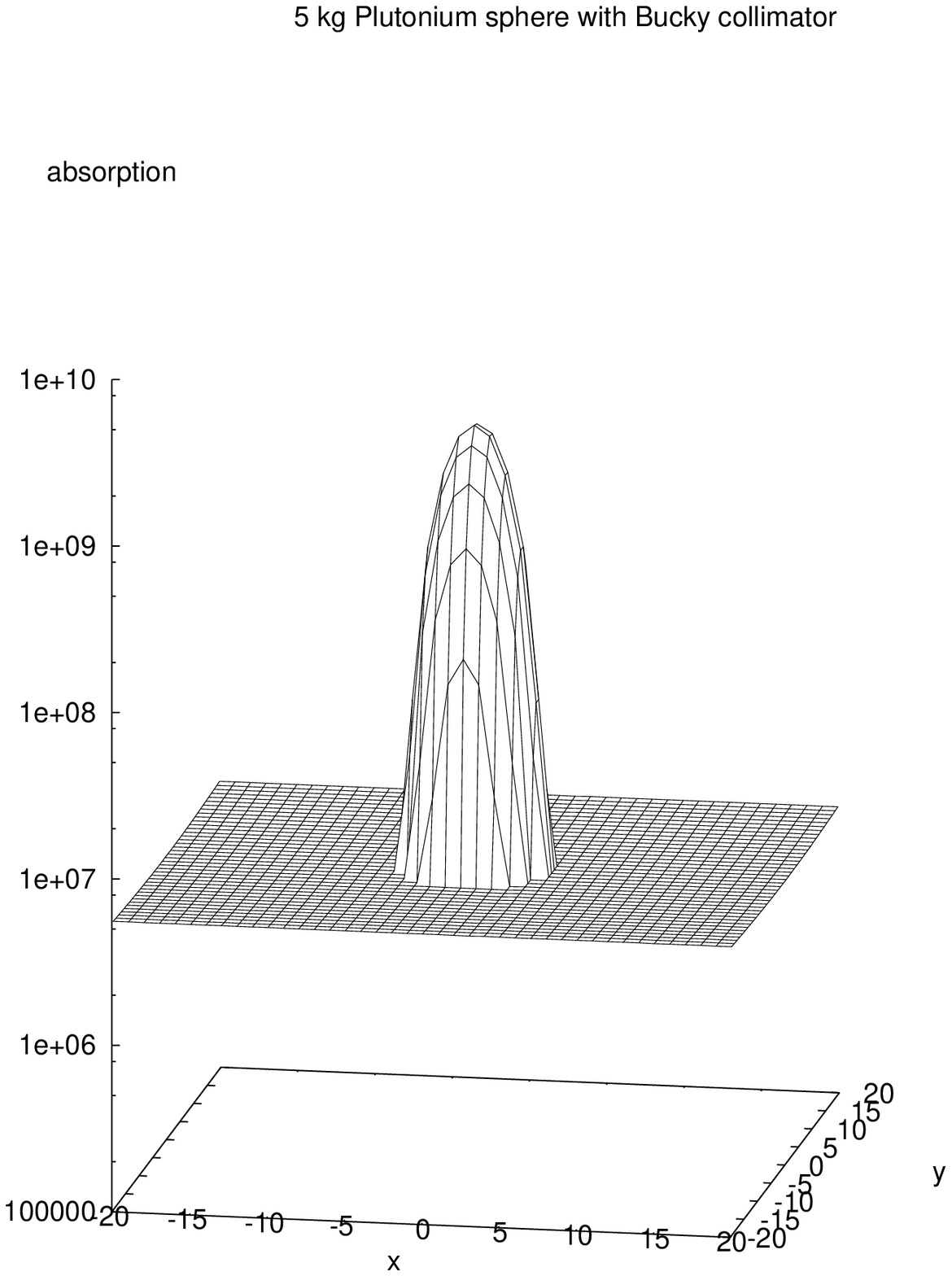}
\end{figure}
\newpage
\begin{figure}
\includegraphics[angle=0,width=5in]{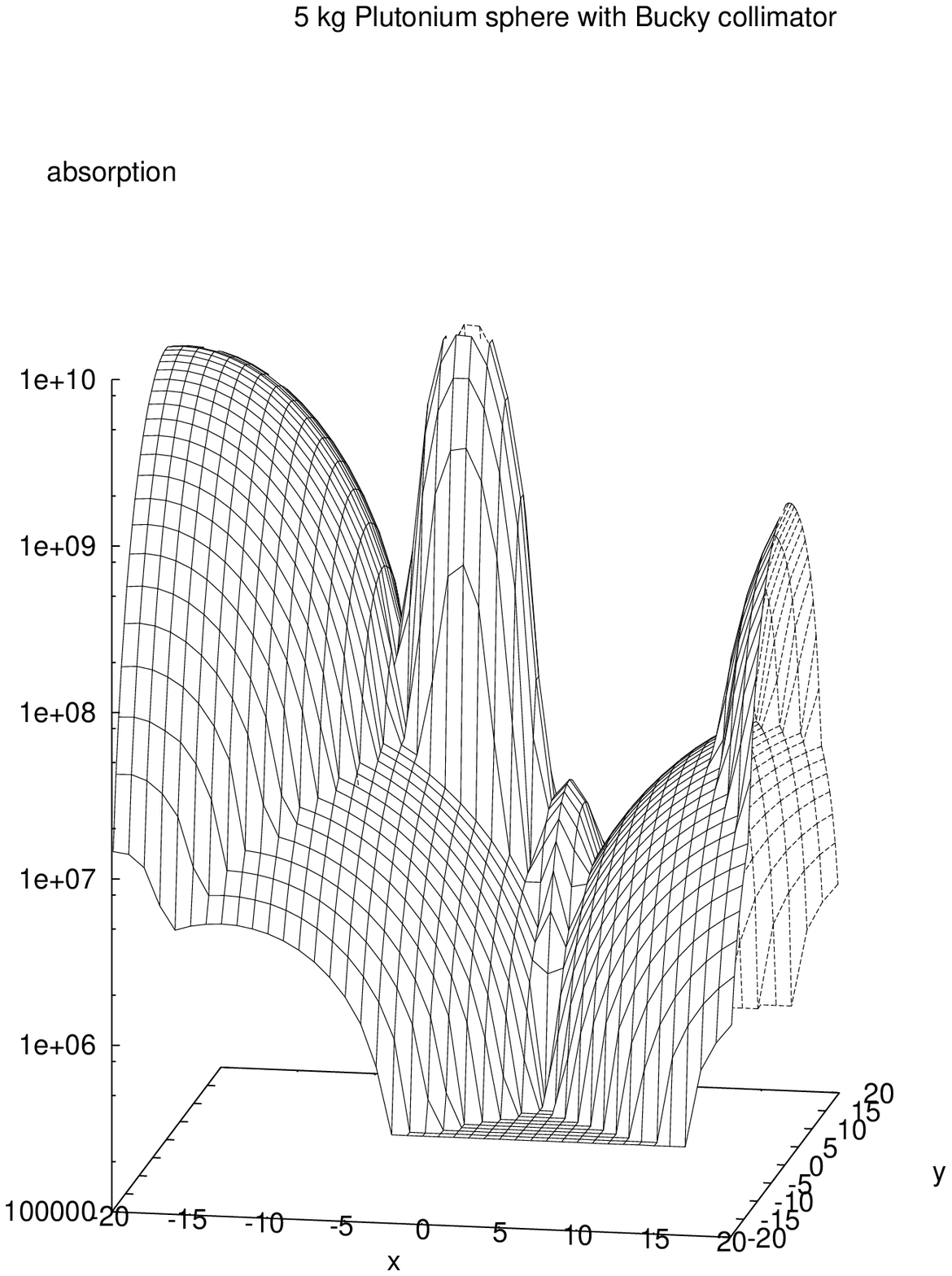}
\end{figure}

\end{document}